%% file: ml-experimental_design.tex
\documentclass[english,round]{article}
\usepackage[T1]{fontenc}
\usepackage[latin9]{inputenc}
\usepackage{geometry}
\usepackage{babel}
\usepackage{float}
\usepackage{amsmath}
\usepackage{amsthm}
\usepackage{amssymb}
\usepackage[authoryear]{natbib}
\usepackage{microtype}
\usepackage[unicode=true,pdfusetitle,
 bookmarks=true,bookmarksnumbered=false,bookmarksopen=false,
 breaklinks=false,pdfborder={0 0 1},backref=false,colorlinks=false]
 {hyperref}

\makeatletter

\floatstyle{ruled}
\newfloat{algorithm}{tbp}{loa}
\providecommand{\algorithmname}{Algorithm}
\floatname{algorithm}{\protect\algorithmname}

\theoremstyle{remark}
\newtheorem{rem}{\protect\remarkname}

\usepackage{iftex}
\DisableLigatures[f]{encoding = *, family = * }

\usepackage{tabularx}
\usepackage{booktabs}

\usepackage{doi}

\usepackage[strict]{changepage}

\ifPDFTeX
  \usepackage{changebar}
\fi

\usepackage{placeins}

\usepackage{pbox}
\usepackage{makecell}
\usepackage{multirow}

\usepackage{booktabs} 

\ifPDFTeX
  \pdftrailerid{} 

  \hypersetup{pdfinfo={ Creator={}, Producer={},ModDate={...},CreationDate={...} }}
  
\else
  \pdfvariable suppressoptionalinfo 767

  \hypersetup{pdfinfo={ Creator={}, Producer={} }}
\fi
\pdfoutput=1
\makeatother

\providecommand{\remarkname}{Remark}

\begin{document}
\title{Machine Learning for Experimental Design: Methods for Improved Blocking}
\author{Brian Quistorff\thanks{Microsoft Technology + Research. Contact: \protect\href{mailto:Brian.Quistorff@microsoft.com}{Brian.Quistorff@microsoft.com}.}
~and Gentry Johnson\thanks{Amazon, AWS Central Economics. Contact: \protect\href{mailto:gentryaj@amazon.com}{gentryaj@amazon.com}.}}
\date{\today}
\maketitle
\begin{abstract}
Restricting randomization in the design of experiments (e.g., using
blocking/stratification, pair-wise matching, or rerandomization) can
improve the treatment-control balance on important covariates and
therefore improve the estimation of the treatment effect, particularly
for small- and medium-sized experiments. Existing guidance on how
to identify these variables and implement the restrictions is incomplete
and conflicting. We identify that differences are mainly due to the
fact that what is important in the pre-treatment data may not translate
to the post-treatment data. We highlight settings where there is sufficient
data to provide clear guidance and outline improved methods to mostly
automate the process using modern machine learning (ML) techniques.
We show in simulations using real-world data, that these methods reduce
both the mean squared error of the estimate (14\%-34\%) and the size
of the standard error (6\%-16\%).
\end{abstract}


\section{Introduction}

In the design of experiments, the method of treatment randomization
can be used to reduce the variance of the estimated treatment effect
so as to improve efficiency, protect against type I errors, and increase
power \citep{Bruhn2009}---particularly for small- and medium-sized
experiments. This is achieved by improving variable \emph{balance}
(similarity of a variable's distribution between the treated and control
groups) for variables that are important predictors of the post-treatment
outcome.

For illustrative purposes we first look at the most common randomization
method, \emph{blocking} (sometimes called \emph{stratifying}\footnote{In some settings ``stratification'' refers to the drawing the experimental
sample from the population and ``blocking'' refers to assigning
treatment.}), originally proposed by \citet{Fisher1935}. Blocking creates a
partition of the sample, separating pre-treatment data into blocks
with a minimum size $c_{B}$ (typically four, \citealt{Ker99}) and
assigning an equal number of treated and control units within each
block. By dividing an important variable's extent with these blocks,
one can increase balance along this variable. For example, if we ensure
that units with different values of an important categorical variable
are partitioned into separate blocks, then we can ensure that, even
in finite-sample (not just in expectation), treatment will be uncorrelated
with this variable.

Existing guidance on how to use pre-treatment data is not fully data-driven
and so involves many decisions by the experimenter, wasting time and
potentially resulting in sub-optimal treatment effect estimation.
There are two main existing strategies for picking blocks. We show
how both can be improved by using modern off-the-shelf machine learning
(ML) solutions that make our proposed procedures mostly data-driven.
We also show how to choose among the available strategies.

The most common approach for determining blocks is what we will refer
to as \emph{variable selection}. This strategy attempts to select
variables that will be strongly related to the post-treatment outcome.
Blocks are then defined by a (potentially uneven) grid from splitting
each variable separately and taking the Cartesian product. The number
of selected variables is kept small as there is a cost to stratifying
on unimportant variables. \citet{Imbens2009} show that while stratification
on a variable can not increase the true expected squared error of
the estimate, it does increase the estimate of its variance when accounting
for stratification due to a degrees-of-freedom adjustment as discussed
below (one could use the variance estimator which ignores the stratification,
but this is overly conservative). In their survey of this approach,
\citet{Bruhn2009} recommend selecting at least the pre-treatment
outcome variable---as most outcomes have some unit-level persistence---and
a geographic variable, as outcome shocks are likely to be correlated
within geographic regions (i.e., the data generating process (DGP)
is time-varying). With multiple pre-treatment periods of data, they
suggest that one could determine which additional variables to include
based on how each early variable is correlated with later pre-treatment
outcomes. Even if there are many related variables, they caution against
including too many since each new variables decreases the balance
on existing ones (given that blocks have a minimum size so that the
granularity along existing dimensions must decrease). In their simulation
studies, they include four blocking variables. They recommend splitting
variables by a roughly even number of quantiles. Overall, the guidance
leaves some areas unspecified (how exactly to determine which variables
to include, how many quantiles to split variables by) and some areas
sub-optimal (partitioning by a grid is sub-optimal when units are
unevenly distributed across multiple selected variables as increasing
the granularity of the grid quickly causes some grid cells to reach
the minimum cell size).

An alternative strategy for blocking has been to use an estimated
prediction model to determine which units to group together. \citet{Barrios2014}
and \citet{Aufenanger2017} both suggest, in different settings, using
pre-treatment data to build a model of the pre-period outcome using
pre-period covariates and then generating predicted values---so-called
pre-period \emph{prognostic scores} \citep{Hansen2008}. Blocks
are formed by ordering units by their prognostic score and then sequentially
allocating blocks of a common size. The guidance here is well specified,
though we note that there may be more optimal ways to partition based
on the prognostic score, given that the default method may create
more blocks than is helpful for minimizing treatment effect error
and therefore result in larger than necessary standard errors.

The approaches differ due to their assumptions about the DGP, particularly
about its temporal properties. If the DGP is constant over time and
well estimated by the predictive model in terms of available predictors
and functional form, then the prognostic score approach is optimal.
It efficiently uses pre-treatment information both by utilizing weakly
related covariates discarded by the variable selection strategy and
also by collapsing all covariates to a single dimension and thus making
it easier to find an optimal partition. Reducing to a single dimension,
however, often results in units with similar prognostic scores that
have very different covariates. If the DGP changes over time, units
with similar pre-treatment prognostic scores may not have similar
future prognostic scores. In this case, it is beneficial to block
instead on a handful of fixed characteristics (e.g., geographic and
demographic variables). Similarly, if the predictive model cannot
closely approximate the functional form of the DGP, it may be more
beneficial to block on separate variables rather than a composite
index.\footnote{For example suppose you estimate a model with a squared covariate
term. Then a unit will be put close to one with the opposite value,
which may differ significantly from the true model.}. If the predictive model is missing variables, then there may be
persistence in the outcome variable over time not captured by the
model, making it beneficial to block on the actual value of the pre-treatment
outcome as this is informative in addition to $\hat{y}_{\textrm{pre}}$. 

We show that, when there are multiple pre-treatment periods of data,
there are ways to choose between the variable selection and prognostic
score strategies. We also apply standard ML tools to automate both.
This includes a strategy for determining the number of blocks, again
an area with little guidance, where we balance the goals of improving
estimate accuracy and reducing standard errors. Most of the improvements
can be made using off-the-shelf ML tools, though we detail some areas
where custom solutions would be helpful.

We note that there are additional situations in which one would want
to block an experiment. It is commonly done if subgroup analysis is
expected to be performed, as (a) the pre-specification guards against
claims of searching indiscriminately for statistically significant
subgroups and (b) it improves the precision of these estimates. We
show how to include these extra block-constraints in the strategies
we present. Finally, it is noteworthy that in the context of two-stage
randomized trials, which we do not study here, one can form blocks
in the second stage (using data from the first stage) to vary the
treatment percentage across blocks in ways that can increase estimation
precision---the so-called \emph{Neyman Allocation} \citep{Tabord-Meehan2018}.

We discuss the basic ML tools involved and our proposed strategies
in Section~\ref{sec:Proposed-algorithm}. In Section~\ref{sec:Extension-Other-randomization}
we discuss the application of these tools to the other most common
methods of randomization. In Section~\ref{sec:Empirics} we use real-world
data to compare our proposed strategies to hand-built blocking. We
conclude in Section~\ref{sec:Conclusion}.

\section{Algorithms\label{sec:Proposed-algorithm}}

We first describe our notation and review basic goals. Then we discuss~a
few standard general ML tasks and highlight the most common method
used currently for each. We then outline our proposed automated strategies
that use these methods and then how to select the optimal one. Finally,
we propose modified strategies when different types of data are available.

\subsection{Econometric setup\label{subsec:Econometric-setup}}

Suppose the following data generating process (DGP)
\begin{align*}
y_{it} & =\beta d_{it}+h_{t}(X_{i})+u_{it}
\end{align*}
where $i\in[1,...,n]$ indexes experimental units (e.g. customers),
$t$ indexes time, and, as above, $d$ is the binary treatment (zero
for all units in the pre-periods and treatment only changes in one
time period), $h$ is potentially time-varying, $X_{i}$ are the observed
covariates, and the $u_{it}$ are independent across units but may
be correlated across time for an individual as we do not measure all
characteristics. We assume we have one post period and at least one
pre-period (i.e. a baseline) of data. We also assume that we will
analyze the experiment using data from $post$ and include dummy variables
for each block. \footnote{Though \citet{Bruhn2009} note that in practice blocking dummies are
often not included, they show empirically that this leads to overly
conservative standard errors. \citet{CPMP} similarly states that
``analysis should reflect the restriction on randomisation implied
by the stratification.''}. 

In this paper, as is widely accepted in the literature on experimentation
and causal inference more broadly, we follow the potential outcomes
framework \citep{rubin1973,holland_1986}. More formally, if we just
consider the $post$ period, then we have outcome $y_{i}$ of unit
$i$ . When unit $i$ receives treatment $d_{i}=1$, her outcome is
$y_{i1}$, and when she receives treatment $d_{i}=0$, her outcome
is $y_{i0}.$ The identification problem is that we cannot observe
unit $i$ in both states, meaning that we must use different units
altogether to serve as the counterfactual for $i$. The average treatment
effect (ATE) is defined as $\beta=E[y_{i1}-y_{i0}]=E[y_{i1}]-[y_{i0}]$.
Without any source of randomization we would estimate

\begin{align*}
\tilde{\beta}= & E[y_{i1}\vert d_{i}=1]-E[y_{i0}\vert d_{i}=0]\\
= & (E[y_{i1}-y_{i0}\vert d_{i}=1])+(E[y_{i0}\vert d_{i}=1]-E[y_{i0}\vert d_{i}=0]).
\end{align*}
The first term on the right-hand side is the the average treatment
effect on the treated. The second is often referred to as selection
bias. To eliminate selection bias the Conditional Independence Assumption,
$\{y_{io,}y_{i1}\}\perp\!\!\!\perp d_{i}\vert X_{i}$, must be met
\citep{rubin_rosenbaum_1983}. In non-experimental causal inference
settings, the intuition underlying this assumption is that, after
controlling for observable characteristics, treatment is as good as
random. It also implies that $E[y_{i1}\vert X_{i},d_{i}=1]-E[y_{i0}\vert X_{i},d_{i}=0]=E[y_{i1}-y_{i0}\vert X_{i}]$,
and therefore $\tilde{\beta}=\beta$.

In experimental settings, it can be easily seen that random assignment
to treatment implies the Conditional Independence Assumption. In fact,
conditioning on $X_{i}$ is not even necessary for the assumption
to hold, and thus inference in experimental settings is generally
far more convincing than in observational settings. While pure randomization
provides identification of $\beta$, more sophisticated treatment
assignment mechanisms such as blocking bring other advantages, as
mentioned briefly above and described in more detail below.

The standard benefits mentioned to motivate blocking include reducing
Type I error, reducing Type II error (increasing power), and increasing
efficiency. Type I error refers to the chance of a false-positive
result given no effect exists. This can happen if there is a finite
sample correlation between the assigned treatment and a prognostic
factor. Blocking will reduce the chance of such a correlation, and
while some experimenters may address this withan ex-post adjustment,
ex-ante restrictions are more efficient \citep{Bruhn2009}.We, therefore,
can reduce Type I errors by reducing the mean-squared error (MSE)
of the estimated treatment effect. Type II error refers to the chance
of failing to detect an effect when one exists. This is directly related
to the variance of the outcomes between the two treatment arms. Blocking
on prognostic factors reduces the sample variances. We, therefore,
can increase power by reducing the standard error of the estimated
treatment effect. Efficiency refers to the number of observations
required to detect an effect for a given experimental setup. While
dependent on many factors, it is typically thought of along the dimension
of power. The more power required, the larger the experiment must
be. We will therefore think about statistical efficiency in terms
of reducing standard errors.

Blocking would, therefore, ideally reduce the estimate's MSE and the
estimate's standard error. These two goals are typically, but not
always, aligned. Blocking on the most important expected prognostic
factors typically improves both, but as mentioned above there is a
degree-of-freedom cost in the estimate's standard error. For example,
using the OLS regression formula, $\widehat{s.e.}(\hat{\beta})=\sqrt{s^{2}(\tilde{X}'\tilde{X})_{11}^{-1}}$,
where $s^{2}=\hat{u}'\hat{u}/(n-b-1)$, $\hat{u}$ are the fitted
residuals, there are $b$ blocks, and $\tilde{X}$ includes $d$ and
along with all the blocking variables. Suppose two blocking partitions,
with $b$ and $b+1$ blocks respectively. As treatment is assigned
orthogonal to the blocking factors, we can ignore differences in the
$(\tilde{X}'\tilde{X})_{11}^{-1}$ term.\footnote{The diagonal elements of $(\tilde{X}'\tilde{X})^{-1}$ measure the
linear dependence of each column of $\tilde{X}$ against the rest.
To see this, let $\hat{x}_{j}$ be the projection of $x_{j}$ on the
subspace spanned by the other vectors, let $\varepsilon_{j}=x_{j}-\hat{x}_{j}$,
and then $(\tilde{X}'\tilde{X})_{jj}^{-1}=1/\left\Vert \varepsilon_{j}\right\Vert ^{2}$
. Since the treatment remains orthogonal its measure of dependence
will remain roughly constant.} The extra blocking may reduce the residuals $\hat{u}'\hat{u}$, but
might increase the standard error through the $\sqrt{1/(n-b-1)}$
term. If the extra blocking does not improve the residuals then the
cost in terms of the relative increase in standard error is 
\begin{align*}
\frac{\widehat{s.e.}(\hat{\tau}_{b+1})}{\widehat{s.e.}(\hat{\tau}_{b})} & =\sqrt{\frac{n-b-1}{n-b-2}}.
\end{align*}
This cost decreases with sample size, and conditional on sample size,
each additional block is increasingly costly (though bounded as the
maximum number of blocks is roughly $n/c_{B}$.) For a sense of scale,
with $n=200$ and $c_{B}=4$, the first partition block increases
the standard error by roughly 0.25\% and the $n/c_{B}$ block increases
the standard error by 0.34\%. For a sample of $n=400$ the numbers
are 0.13\% and 0.14\%. The partition, therefore, that minimizes
the estimate's standard error can have fewer blocks than the one that
minimizes the estimate's MSE. If any block in the partition that
minimizes the effect standard error has a size of at least $2c_{B}$
then splitting it would improve the estimate's MSE.

An additional motivation for blocking can emerge if the experimenter
expects to perform subgroup analysis across a particular variable
to look for heterogeneity. In this case, the experimenter would have
selected $\tilde{X}$ variables with an existing partition (likely
a grid) $\tilde{\Pi}$ for this analysis. For example, in the simplest
case, all $\tilde{X}$ variables could be split at their median values
creating $2^{|\tilde{X}|}$ initial blocks. In Section \ref{subsec:Strategy:-Variable-selection}
and Section \ref{subsec:Strategy:-Future-Prognostic}, we suggest
minor adaptations to the procedures in this paper when an ex-ante
subgroup analysis plan motivates blocking.

Given the two goals (of reducing the estimate's MSE and standard error)
might diverge, one must decide on an overall strategy. Ex-post one
could optimally select among candidate partitions given a weighting
between the two goals, but ex-ante this is much harder. While we note
later how to do this when data is available (see Section~\ref{subsec:3+-pre-periods}),
typically the data will not be available or can be put to better use.
We, therefore, put forward a simple and reasonable approach that tries
to balance them without explicitly optimizing the two goals. We create
a sequence of partitions (discussed below) and find the partition
where its fitted outcome model (fitting $y$ to the block variables)
has the best expected out-of-sample accuracy. This naturally limits
the partition complexity to some degree, as when building a predictive
model, a partition that is too fine-grained will over-fit to its training
data and perform badly out-of-sample (e.g., a partition with a block
for every observation has clearly gone beyond finding generalizable
patterns and instead memorizes the idiosyncrasies of the current sample).
This strategy, of estimating the out-of-sample performance of a model,
can be done efficiently using the same data via a procedure called
\emph{cross-validation} (discussed below).

The value of blocking decreases with sample size. As the sample size
increases, the chance of a finite-sample correlation between treatment
and a prognostic factor decreases. Therefore the concern about the
estimate's MSE becomes less important. For statistical efficiency,
it is commonly believed that blocking in larger samples is less important
\citep{Ker99}, but this will depend more on the nature of the data.
Overall, many clinical trialists suggest blocking is less important
with samples over 400 \citep{Ker99}. Our view, however, is that if
the process can be made easy, then the benefits may outweigh the costs
at many sample sizes.

We focus first on the situation of having at least two pre-periods
worth of data $t\in\{\textrm{pre1},\textrm{pre2}\}$ as this is the
cleanest setup for the models. Both strategies will model the relation
between $y_{\textrm{pre2}}$ and $[X,y_{\textrm{pre1}}${]} to form
partitions, and we show how to use an out-of-sample method to pick
between them. After deciding which strategy to use, given there is
likely some temporal dependence, we proceed with estimation by using
the selected model to generate partitions when using $[X,y_{\textrm{pre2}}]$,
rather than $[X,y_{\textrm{pre1}}]$.

\subsection{Standard ML methods}

Before we detail the ML methods, we first discuss a general difference
with more common methods used in economics. ML models typically have
\emph{hyper-parameters, }which often control the model's overall complexity.
One benefit of many ML methods is that they can be quite complex,
but increasing their complexity too much can mean that they \emph{overfit}
to the sample data, essentially memorizing the idiosyncrasies of the
current sample and behaving badly out of sample. Experimenters and
practitioners have therefore developed procedures to modulate model
complexity and limit overfitting. The main procedure is \emph{cross-validation}
(CV), which simulates the out-of-sample error. CV randomly split the
data into $K$ ``folds'' (usually 5 or 10). \emph{Out-of-sample}
predictions are made for each observation using a model that was trained
on all data but the fold for that observation (so there are $K$ separately
trained sub-models). One can then fit the model with different values
of the hyper-parameter and pick the hyper-parameter value based on
the one with the lowest mean squared prediction error (MSPE).\footnote{An alternative is the ``1se'' rule \citep{Friedman2010}, which
is the simplest model that has a MSE no more than one standard error
above the minimum. This is typically used if there is a strong reason
to believe that the model will be used with new data drawn from a
different distribution.}

We list three common ML tasks and identify for each a method that
is common, simple, and can be used off-the-shelf :
\begin{itemize}
\item Partitioning: This task is to create a partition, $\Pi$, with cells
$\ell$, from a feature-space $X$ to a variable level of complexity.
The goal when constructing the partition is that the set of dummy
variables covering each block maximizes their predictive power for
$y$ (i.e., the predicted value for each block is the block's mean
outcome). Finding the globally optimal partition is too computationally
intensive, so the most common method \citep{Hastie2009} for this
task is Classification and Regression Tree (Cart, \citealt{Breiman1993}).
Cart starts with the whole feature space as a single block and recursively
splits each block into two using rectilinear cuts. To split a block,
it searches over each dimension and possible values in that block
and finds the split that reduces the overall MSE of the outcome of
the two sub-blocks. Intuitively, it finds a split such that the two
sides have very different mean outcomes. The main hyperparameters
are the tree depth (which we choose through CV) and the minimum leaf
size (which we set as $c_{B}$)
\item Feature selection: In this task, we have a generic outcome $y$ and
features $X$ and we would like to find the subset $X^{*}$ that is
most important for determining $y$. The most common method \citep{Taddy2019}
is the Least Absolute Shrinkage and Selection Operator (Lasso, \citealt{Tibshirani1996}).
Lasso is a linear model that adds to the OLS objective function a
penalty for the $L_{1}$-norm of the coefficients, solving $\min_{\beta}||y-X\beta||_{2}^{2}+\lambda||\beta||_{1}$.
The Lasso solution will typically set many coefficients to exactly
zero due to the geometry of the $L_{1}$ penalization. If the true
DGP is sparse in terms of the non-zero coefficients, then under certain
conditions the Lasso can achieve the \emph{oracle property} and be
consistent in terms of selecting the true subset \citep{Zou2006}.
We highlight three usage notes. First, as the absolute sizes of the
coefficients are all penalized, we typically normalize all features
to have a standard mean and variance. Second, as it's a linear model,
variables that interact or affect the outcome non-linearly may not
be selected.\footnote{The partitioning methods introduced above may also be used as a non-linear
variable-selection procedure by generating a partition and then selecting
the variables that were used to split on at least once. With many
variables, however, the performance of decision trees suffers \citep{Hastie2009}
and so in these cases especially, Lasso is preferred.} To help address this, a common practice is to augment $X$ with common
transformations. Third, we will follow common practice and set the
$\lambda$ hyperparameter using CV.\footnote{We note that for Lasso, some plug-in estimates for setting $\lambda$
have attractive theoretical properties \citep{Belloni2012}.}
\begin{itemize}
\item A common sub-task is to identify importance weights $\{w_{k}\}$ of
the selected variables. As the Lasso coefficients are biased due to
the $L_{1}$ regularization, one can construct importance weights
by performing a subsequent OLS on just the Lasso-selected features
(the Post-Lasso by \citealt{Belloni2013b}) and taking the absolute
value of the coefficients.\footnote{For more complicated methods, one can re-run the models each time
omitted one covariate and use the increase in MSE of the outcome as
a measure of importance.}
\end{itemize}
\item Prediction: In this task we wish to form a robust prediction in the
face of potential non-linearities, learning $y\approx\hat{g}(X)$.
There are many options for this task, but in most statistical data
(i.e., not visual or text data) applications, the Random Forest \citep{Breiman2001}
is common, simple, and performs well \citep{Taddy2019}. A Random
Forest is the average of a large number of separate tree models (typically
Cart). Each tree is trained on a slight modification of the original
data (the data is bootstrapped and then at each splitting decision
a random number of features are selected as candidates for splitting),
to yield different trees adding smoothness and robustness. 
\end{itemize}
We note that, while we have picked a popular, widely available, and
simple method for each purpose, there are alternatives (e.g., Best
Subset instead of Lasso and Boosted Trees instead of Random Forests).
If there are data or computational reasons to pick an alternative,
that should be explored by the experimenter. The above can be thought
of as default choices to operationalize the below algorithms. In the
description of the algorithms, we will use the generic task name---partitioning,
feature selection, or prediction---rather than any particular method.

Finally, as with most ML methods, the ones noted here can function
even when there are more features than observations. They are therefore
quite useful in settings where, despite a small sample size, we nonetheless
have rich data on individuals.

\subsection{Strategy: Variable selection (VS)\label{subsec:Strategy:-Variable-selection}}

As mentioned above, we can use a dedicated feature selection method
initially or directly use the partitioning algorithm. The choice will
depend on the number of covariates, $K$, and the experimenter's prior
on the sparsity of the covariates in the DGP. If $K$ is relatively
small, then the partitioning algorithm can be used directly on the
variables to create blocks. If $K$ is relatively large, then the
performance of partitioning algorithms tends to suffer. In that case,
and especially if the experimenter's prior is that only a sparse subset
of the variables matter for predicting the outcome, we can use a preliminary
feature selection method.

Note that in addition to the standard variables, we could pre-generate
$\hat{y}_{\textrm{pre1}}$ (from a prediction model of $y_{\textrm{pre1}}\approx g_{PS}^{\textrm{pre1}}(X)$)
and include it as well. Its inclusion might improve performance and
focuses this strategy on alleviating issues rising from model misspecification
and dynamic DGPs. Performance improvements will result if there is
a long tail of covariates in $X$ that are weakly related to $y_{pre1}$
and can therefore be compactly represented in $\hat{y}_{pre1}$. An
orientation towards the issue of dynamic DGPs will result insofar
as $\hat{y}_{\textrm{pre1}}$ captures information from $X$ that
explains the static components of the DGP, meaning the covariates
selected from $X$ when $\hat{y}_{\textrm{pre1}}$ is included in
the model will be those whose influence may vary over time. Concisely,
if $y_{\textrm{pre1}}$ is selected, this is evidence of persistence
(unspecified variables), and if some of $X$ is selected, then this
is evidence of a dynamic DGP.

In some circumstances, blocking on real variables (even if chosen
by a model) may be preferred for interpretability and trustworthiness
reasons to using a synthetic feature such as $\hat{y}_{\textrm{pre1}}$.
Using a synthetic measure such as these can result in \emph{unintuitive}
groups (units that, while having similar prognostic scores, have very
different covariates). This concern is raised similarly in the matching
literature \citep{King2016} in the context of propensity-score matching.
If interpretable blocks are required, then the experimenter may prefer
to leave $\hat{y}_{\textrm{pre1}}$ out of the Variable Selection
strategy.

Full details are in Algorithm~\ref{alg:Variable-Selection-Blocking-Algo}.

\begin{algorithm}
\caption{Variable Selection Blocking Strategy\label{alg:Variable-Selection-Blocking-Algo}}

\emph{Inputs}: $y_{\textrm{pre1}},y_{\textrm{pre2}}$, and $X$.
\begin{enumerate}
\item Estimate a prediction model $y_{pre1}\approx g_{PS}^{\textrm{pre1}}(X)$
and generate $\hat{y}_{\textrm{pre1}}$. Define $M=\{y_{\textrm{pre}1},\hat{y}_{\textrm{pre1}},X\}$.
\item Estimate a prediction model $y_{pre2}\approx g_{PS}^{\textrm{pre2}}(X)$
and generate $\hat{y}_{\textrm{pre2}}$.
\item If $K$ is large (or assuming sparsity): Use a feature selection method
predicting $y_{\textrm{pre2}}$ using $M$. Redefine $M$ as the selected
set of features. (If needed for a downstream task, return the importance
weights).
\item Perform partition (with CV tree depth) predicting $y_{\textrm{pre2}}$
using $M$, yielding partition $\Pi$.
\item Assign blocks based on updated data: $b=\Pi(y_{\textrm{pre2}},\hat{y}_{\textrm{pre2}},X)$.
Ensure that the partition did not create blocks smaller than $c_{B}$
with updated data (if so, prune back the tree complexity until this
constraint is satisfied).
\end{enumerate}
\emph{Return}: $b$
\end{algorithm}

This algorithm has multiple advantages over the existing manual process:
\begin{enumerate}
\item This algorithm has a common method for selecting among $y_{\textrm{pre1}}$,
geographic variables, and other features. It also focuses on predictive
power in a joint setting rather than using bivariate correlations.
While the feature selection method does not jointly pick blocking
variables and partition, it does have an automatic stopping rule (the
cross-validated $\lambda$) for limiting the selected set of blocking
variables.
\item In general, a tree-based partition is preferred to a grid-partition
as it can have increased granularity while adapting to densely and
sparsely populated regions of the covariate space. \footnote{Technically the minimum blocks size is only used on the $\textrm{pre2}$
data, but blocks could be made smaller for the $\textrm{pre1}$ data
as well. In practice, typical decision trees have minimum node leaf
sizes of around 6 so as to not estimate means from very small samples.
As a result, this variant is unlikely to is be helpful.}
\item As the partition algorithm ensures that there is expected benefit
to a finer partition, we naturally balance the trade-off between increased
granularity and the downstream degrees-of-freedom adjustment.
\end{enumerate}
If the experimenter is unsure whether to use an initial variable selection
method in Algorithm~\ref{alg:Variable-Selection-Blocking-Algo},
one can create both versions of the variable selection strategy and
use the procedure in Section~\ref{subsec:Deciding-between-FPS} to
decide between them. If the experimenter is motivated to block in
order to carry out a pre-specified subgroup analysis, then we suggest
the following modification to Algorithm \ref{alg:Variable-Selection-Blocking-Algo}:
in the partitioning step, we start with the existing partition $\tilde{\Pi}$
as described in Section \ref{subsec:Econometric-setup} and recursively
partition cells from that point. If the experimenter is using an initial
feature selection, then the procedure should be constrained to only
allow new splits on selected dimensions.
\begin{rem}
\emph{\label{rem:Adaptive-Grid-Alternative}(Adaptive Grid Alternative)}
The partition created by Cart will subdivide the space into hyperrectangles,
but the partition can still be quite irregular and hard to understand.
If the partition needs to be understandable on its own, then an alternative
is to use an \emph{adaptive grid partition}. This grid can be built
by dividing (when possible) covariates on quantiles. There should
be more blocks across variables that are more important. Therefore,
attempt to make the number of blocks across variables roughly proportional
to their importance weights. The overall granularity is a hyperparameter
that can be set by CV (and assuming a minimum block size of $c_{B}$).
\end{rem}
\begin{rem}
\emph{(Misfits)} Blocks for experiments often contain an odd number
of units, preventing a perfect even distribution of treatment in the
block. One of the units (typically at random) is held-out and labeled
the ``misfit'' and the rest are randomized event across treatments.
One may want to ensure that the treatment assignment of the misfits
is also even across the distribution. If blocks span only a single
dimension then we can iterate across the blocks in order and assign
misfits to alternating treatments. If blocks span multiple dimensions,
however, there are no simple solutions. (If the misfits themselves
form a rectangular lattice then this is possible, but this is highly
unlikely). Practice varies in this situation and non-random solutions
are typically slow and approximate. 

One approach, with any progressive partition method, is to view just
the misfit units from a higher, coarser-level of partitioning and
re-do blocking at this higher level. With Cart, this can be done easily
by simply iterating across the tree leaves in order and assigning
misfits to alternating treatments. Subsequent misfits will come from
the same part of the original partition and would therefore be in
the same cell at a higher level.
\end{rem}
\begin{rem}
\emph{(Feature learning)} In ML, a related task to feature selection
is \emph{feature learning}. This focuses on generating (often a small
set of) synthetic features that are transformations or combinations
of the original features that can perform better than the original
features for some downstream estimation. This is a task that many
experimenters already do manually (e.g., constructing composite indexes,
averages, and log/polynomial transformations of existing features),
but feature learning performs this in an automated way. Learned features
are often constructed using neural networks \citep{Hinton504}. A
full treatment of the theory and application of feature learning is
beyond the scope of this paper, so we note here merely situations
where feature learning might be helpful. It would be a case where
FPS will not perform the best (the true DGP is difficult to approximate
or the DGP is dynamic), but where VS does not perform as well as should
be expected (e.g., because there are too many variables to select,
so some combination is helpful). We note that given the procedure
must learn an additional set of transformations, the task usually
requires a larger sample size, potentially limiting its usefulness.
To our knowledge, feature learning has not yet been applied to experimental
blocking.
\end{rem}

\subsection{Strategy: Future Prognostic Score (FPS)\label{subsec:Strategy:-Future-Prognostic}}

We construct a \emph{Future Prognostic Score (FPS)} by using a prediction
model to approximate\footnote{A pre-generated $\hat{y}_{\textrm{pre1}}$ could be used here as well,
but it would not improve performance unless it were estimated using
a different algorithm.}
\begin{align}
y_{\textrm{pre2}} & \approx g_{FPS}(X,y_{\textrm{pre1}}).\label{eq:Ypre2_using_pre1}
\end{align}
Note that this is different from the simple prognostic score models
of \citet{Barrios2014} and \citet{Aufenanger2017}, as it is looking
one step ahead and incorporates a past outcome value. This ensures
that this strategy uses the same data as the variable selection strategy.
With the past outcome value, FPS can now deal with outcome persistence,
though since it collapses the match-space to a single index it can
not deal with a dynamic DGP. We must still also consider the fact
that our model may be misspecified.

As with the above, blocking is carried out on the predicted value
using updated features $\hat{g}_{FPS}(X,y_{\textrm{pre2}})$. The
existing standard method, which we call \emph{Sequential Allocation,}
is to arrange units according to their FPS and generate groups of
size $c_{B}$. Groups can be made larger to incorporate segments of
units with identical predicted values. This ensures that extra blocks
are only created when there is a benefit to (in-sample) predictive
performance. This might create more odd-sized cells, but misfits are
less of a problem in this approach as we can ensure an even distribution
of the treatment arms across the span of the prognostic scores by
iterating across the misfits in order and alternately assigning treatment.In
the case of blocking motivated by a pre-planned subgroup analysis,
the experimenter should start with the existing partition $\tilde{\Pi}$
as described in Section \ref{subsec:Econometric-setup}, arrange units
within each block by their FPS, and procced partitioning from that
point (ensuring no cell with size below $c_{B}$). 
\begin{rem}
\label{rem:Alternate-score-based-partition}(Alternate score-based
partitioning) The existing approach of taking prognostic scores and
performing Sequential Allocation may create too many blocks since
it focuses on in-sample predictive performance. The first-stage predictive
method for learning $\hat{g}_{FPS}$ does use tools to control for
over-fitting (so that $\hat{y}_{i}$ is not too influenced by $y_{i}$),
but will likely still create too many unique levels of $\hat{y}$
and that is all the Sequential Allocator focuses on. We need to treat
the joint process of learning $\hat{g}_{FPS}$ and constructing the
allocation as a combined partition method using CV to control for
the final complexity (the number of blocks). Given we want a second-stage
partitioning method that can create a partition with a less-than-maximal
number of blocks, we may want more complexity than the Sequential
Allocator. Options include:
\begin{itemize}
\item Simple: A Scaled Sequential Allocator that creates fewer blocks than
$N/c_{B}$, but still roughly evenly sized. This is simple, but far
from optimal.\footnote{We note that another simple alternative would be to use Cart targeting
$y_{\textrm{pre2}}$ and blocking on $\hat{y}_{\textrm{pre2}}=\hat{g}_{FPS}(X,y_{\textrm{pre1}})$
to create the partition. This solution likely does not offer any
benefit in single-dimensional partitioning as the greedy solution
will results in a very uneven distribution of sizes (some blocks roughly
twice the size of others). }
\item Complex: Since we are only dealing with a single dimension, there
will be many fewer possible partitions, and we can jointly optimize
the splitting rules rather than use a greedy solution such as Cart.
A straightforward approach would start with quantile splits and then
use coordinate-descent to sequentially optimize each split until no
changes are made.
\end{itemize}
Regardless of the actual partitioning method used, the complexity
should still be tuned for CV performance. As this is a two-stage process,
for each iteration $f$ we learn a separate $\hat{g}_{FPS}^{f}$ and
partition using all data but fold $f$ creating an outcome prediction
of the average prognostic score in each block and then see the out-of-sample
performance on fold $f$.
\end{rem}

\subsection{Deciding between FPS and VS\label{subsec:Deciding-between-FPS}}

There are different ways to determine which strategy to use depending
on the available data:
\begin{itemize}
\item If there is another pre-treatment period, $\textrm{pre3}$, we can
empirically see which resulting partition has the best predictive
performance on$y_{\textrm{pre3}}$.
\item If not, then we can compare performance using cross-validation, where
here we choose between different model types rather than between different
hyper-parameters for a single model type.\footnote{Note that comparing directly the performance of the partitions from
the above models on $y_{\textrm{pre2}}$ will be biased as the ML
models were trained on that data.}. Given we need to have sufficient units per block, a 2-fold CV version
is best to maximize the size of the held-out fold. One can average
the results over multiple random splits to reduce noise. Note that
this works best with larger datasets.
\end{itemize}
After deciding which strategy to use, given there is likely temporal
dependence, we use the model to generate partitions when using $y_{\textrm{pre2}}$
rather than $y_{\textrm{pre1}}$.

\subsection{Different pre-period data}

\subsubsection{Time-varying covariates}

If there are time-varying covariates $Z_{it}$, then they should be
used in the same way as $y_{\textrm{pre}}$. $Z_{\textrm{pre1}}$
would be used when modeling $y_{\textrm{pre2}}$, and then updated
values $Z_{\textrm{pre2}}$ would be used to construct the final partition.

\subsubsection{3+ pre-periods\label{subsec:3+-pre-periods}}

With more time-periods, we can improve several parts of the process.
One option is to use the above strategies with additional look-ahead
predictions.
\begin{itemize}
\item Variable selection: Use variables $M=\{y_{\textrm{pre2}},y_{\textrm{pre1}},X,\hat{y}_{\textrm{pre1}},\hat{y}_{\textrm{pre2}}\}$
and either construct the partition directly or via an initial feature
selection method targeting $y_{\textrm{pre3}}$.
\item FPS: Generate prediction values from $y_{\textrm{pre3}}\approx g_{FPS+}(y_{\textrm{pre2}},y_{\textrm{pre1}},X)$
\end{itemize}
A second option is to use $y_{\textrm{pre3}}$ to find an optimal
trade-off between the goal of reducing the estimate's MSE and its
standard error. For every candidate partition created using data from
$\{y_{\textrm{pre2}},y_{\textrm{pre1}},X\}$, one could simulate $S$
different randomizations and then calculate the average standard error
and MSE (given there was no actual treatment in $\textrm{pre3}$).
Given an optimal weight between the two goals, the best partition
could be selected. 

When data is scarce, using the extra information to inform the ML
models is likely more beneficial. If there are strong reasons to believe
that the default partition complexity is non-optimal (e.g., idiosyncratic
research preferences) then the latter option may be preferred. 

\section{Other randomization methods\label{sec:Extension-Other-randomization}}

We review here the other main randomization methods, \emph{pair-wise
matching} and \emph{rerandomization}, and how the above strategies
can be modified when either are preferred to blocking.

\subsection{Pair-wise matching}

Pair-wise matching divides the sample into similar pair with each
pair randomly assigned to have a treated and control unit. If the
experimenter wants to improve balance along a certain variable, this
can be explicitly achieved through including that variable in the
match criteria.

Application of Strategies:
\begin{itemize}
\item Variable selection: It is straightforward to use the feature selection
method above to select a match space and then construct pairs. Each
unit has values for its selected features, $M$, and so our task becomes
to divide the units into pairs with a method that attempts to minimize
the overall within-pair differences (where we define distance as geometric
distance in $M$, but where we weight each dimension by its importance
$w_{k}$). This is similar to the problem of matching treated to control
units in 1-1 matching estimators. Similar to that domain, the optimal
solution \citep{Greevy2004} is quite difficult, so most implementations
take the approach of finding the ``nearest available match'' \citep{King2007}.
We, therefore, suggest the same: select available units randomly and
pair them to their nearest available unit.
\item Future prognostic score: Use a prediction model to generate prognostic
scores, order units by their score, and then sequentially put them
into pairs.
\end{itemize}
Selection between strategies: As we can produce pair-level dummies
similar to the block-level dummies, the selection procedure is the
same as with blocking.

\subsection{Rerandomization/Minimization}

Rerandomization techniques \citep{Taves1974,Pocock1975} repeatedly
randomize units to treatment and control arms until the imbalance
across important variables meets some criterion. There are two methods
commonly used: ``big stick'' which rerandomizes until no important
variable has a significant difference at a pre-specified level (commonly
5\%) and ``min-max'' which computes for a pre-specified $R$ number
of draws (commonly 1000) the maximum $t$-statistic difference for
the important variables and then chooses the randomization with the
minimum maximum. Notice that, in contrast to the other methods, this
ensures a parametric rather than non-parametric form of balance as
we explicitly specify the moments (typically means) that should be
matched. We will focus on the min-max strategy, but it is straightforward
to adapt the methods for the ``big stick'' approach. Let $\theta_{rk}$
be the $t$-statistic for the difference in means of the $k$th variable
between the two treatment arms in the $r$th randomization, so that
the standard min-max strategy selects $r^{*}=\arg\min_{r}[\max_{k}\theta_{rk}]$.

Application of strategies:
\begin{itemize}
\item Variable Selection: Proceeding as we have above in the variable selection
setting, we use the feature selection method as to get selected variables
$M$. This will constitute the set of variables for which we will
compare the t-statistics of mean differences across treated and control
units. We suggest taking into account the relative importance of the
variables by finding the ideal randomization via $r^{*}=\arg\min_{r}[\max_{k}w_{k}\theta_{rk}]$.
\item Future Prognostic Score: Use a prediction model to generate future
prognostic scores. Let $\tilde{\theta}_{r}$ be the $t$-statistic
for the difference in means of the future prognostic scores for the
$r$th randomization. As we have collapsed the dimensions we now simply
choose $r^{*}=\arg\min_{r}\tilde{\theta}_{r}$.
\end{itemize}
Selection between strategies: If we have access to an additional pre-period
of data, then we can choose between the above methods in a similar
way by taking both approaches and seeing how well they do at minimizing
average differences in $y_{\textrm{pre3}}$ between treatment and
control groups. If we do not, we can use the method for blocking using
CV and see the average difference between the arms in the hold-out
samples.

\section{Simulations\label{sec:Empirics}}

To analyze empirically how well our strategies perform, we use the
data and framework of \citet{Bruhn2009}, comparing their manually
constructed blocks against our blocking strategies. We use the two
datasets from their framework containing more than two pre-treatment
outcomes periods: a panel survey of microenterprises in Sri Lanka
\citep{Mel2008} and a sub-sample of the Mexican employment survey
(ENE). In both of these, the subgroup studied received no treatments.
We treat the first two periods as $\textrm{pre1}$ and $\textrm{pre2}$
and the third as $\textrm{post}$. For both, we estimate results using
the n=100 and n=300 samples. The Sri Lanka dataset has 29 covariates
and the Mexican sample has 30 covariates. The benefit of the ML strategies
we propose typically increases with the number of covariates. We perform
10,000 simulations of placebo assignments to units and assess the
performance of the strategies above as compared to the strategy of
\citet{Bruhn2009} that constructs 48 blocks by hand-picking four
variables and then manually determining a grid. 

We analyze our results in terms of MSE of the treatment effect (given
we know the true effect is zero) and the size of the standard error.
Table~\ref{tab:Coefficient-MSE} reports the MSE of the estimated
coefficient. We see that all of our strategies perform better than
the manual method across all samples. The reduction in the MSE from
using the best ML method ranges from 16\%-34\%. The Future Prognostic
Score strategy performed best on the Mexican ENE sample with N = 100,
sample whereas the Variable Selection strategy with initial Feature
Selection performed best on the Mexican ENE sample with $N=300$ and
both Sri Lankan samples.

Table~\ref{tab:SE_length} reports the length of the standard error
for the estimate, a measure of increased precision. All ML algorithms
again perform better than the manual strategy across all samples.
The reduction in the MSE from using the best ML method ranges from
6\%-16\%. All three automated strategies performed best in at least
one context.

\begin{table}
\caption{Coefficient MSE\label{tab:Coefficient-MSE}}

\input{insets/tbl2_coeff_controls2.tex}
\end{table}

\begin{table}
\caption{Size of Coefficient Standard Error\label{tab:SE_length}}

\input{insets/tbl2_se_controls.tex}
\end{table}

\section{Conclusion\label{sec:Conclusion}}

Restricting randomization in experiments to reduce treatment-control
imbalances on variables that are important for predicting the post-treatment
outcome improves efficiency, protects against type I errors, and increases
power for the estimated treatment effect \citep{Bruhn2009}, particularly
for small- and medium-sized samples. Existing guidance for this process
has been conflicting and demands many \emph{ad hoc} decisions. We
show that this incompleteness in guidance is due to differing views
on the dynamics of the data generating process (DGP). In the case
of having at least two pre-periods worth of baseline data, we outline
methods that resolve these differences and automate the process using
modern, and off-the-shelf machine learning (ML) techniques. For the
main type of randomization restriction, blocking, we determine what
are the important dimensions to create blocks along, how to create
blocks, and how many should be made. Crucially, for determining how
many blocks to create, we provide a way to balance the goal of improving
the estimators true accuracy, which improves with more blocks, and
the goal of reducing the estimated standard error, which can increase
due to a degree-of-freedom correction if the extra blocks are only
marginally helpful. Applications are also show to the other main types
of randomization restrictions: pair-wise matching and rerandomization.
With real-world data, we see reductions in the mean squared error
of the estimated coefficient of 14\%-34\% and reductions in the standard
error of the estimate of 6\%-16\%. We also detail custom tools that
may improve performance even more.

\bibliographystyle{plainnat}
\bibliography{matching}

\appendix

\section{Alternative tree splitting rule for partitions\label{sec:Alternative-tree-splitting}}

While partitioning algorithms are fit on one set of data, they are
designed to not overfit to the sample and are instead tuned to do
well on the general population of data that the sample was drawn from.
The standard way to do this is to fit a full sequence of partitions
of increasing granularity, each focusing on in-sample fit, and then
to pick the one that does best on CV OOS predictions. An alternative
way, pioneered by \citet{Athey2016a}, is to incorporate this out-of-sample
focus directly into each splitting decision in cases where we know
the size of the auxiliary sample on which the partition will be used.
Taking the example of Cart, one can write the typical objective function
as finding the partition $\Pi$ that minimizes the ``modified''
MSE
\begin{align*}
\textrm{MSE}(\Pi;\mathcal{S}^{\mathrm{pre}}) & =-\frac{1}{N}\sum_{\ell\in\Pi}N_{\ell}\hat{\mu}^{2}(\ell;\mathcal{S}^{\mathrm{pre}},\Pi).
\end{align*}
\citet{Athey2016a} show that if we take the auxiliary sample into
account during the split we should minimize the Expected MSE, which
can be estimated as
\begin{align*}
\mathrm{\widehat{EMSE}}(\Pi;\mathcal{S}^{\mathrm{pre}}) & =-\frac{1}{N}\sum_{\ell\in\Pi}N_{\ell}\hat{\mu}^{2}(\ell;\mathcal{S}^{\mathrm{pre}},\Pi)+\frac{2}{N}\sum_{\ell\in\Pi}N_{\ell}\hat{\mathbb{V}}(\hat{\mu}(\ell;\mathcal{S}^{\mathrm{pre}},\Pi))
\end{align*}
where we now penalize blocks that have high variance in their estimates.
Using this for partitioning requires custom tools (\citet{Athey2016a}
provide tools for partitioning on estimated treatment effect, not
estimated outcome), so we leave this for future work. 

\section{Alternative available data}

\subsection{1 pre-period}

This is the typical case studied in the previous literature. We can
automate a few portions of the standard strategies, but we can not
deal with the general temporal dynamics of the DGP:
\begin{itemize}
\item Variable selection: Since we only have a single outcome, we do not
have a separate target to jointly pick the best variables from $[X,y_{\textrm{pre1}}]$.
We, therefore, take the guidance of \citet{Bruhn2009} and force the
inclusion of $y_{\textrm{pre1}}$ and separately select the features
$X^{*}$ from a feature selection model targeting $y_{\textrm{pre1}}$
with $X$. Similarly, we no longer can construct a partition based
on a joint predictive model. We could construct an \emph{adaptive
grid} (as above). The experimenter would have to give a relative weight
for $y_{\textrm{pre1}}$ compared to the variables in $X^{*}$. Obvious
candidates would be $\sum_{k\in X^{*}}w_{k}$ (so that $y_{\textrm{pre1}}$
has equal weight to all of $X^{*}$) or $\frac{1}{|X^{*}|}\sum_{k\in X^{*}}w_{k}$
(the average weight from $X^{*}$).
\item Prognostic score: Construct the simple prognostic scores from a model
of $y_{\textrm{pre1}}\approx g_{PS}(X)$. Then order units by their
prognostic score and partition them into groups of $c_{B}$.
\end{itemize}
Selection between strategies: Here the experimenter would have to
take a stand on the amount of temporal dependence in the DGP (which
could potentially be assessed in another data source).
\begin{rem}
(Auxiliary sample) If there is an auxiliary sample with improved data
(e.g., $[X,y_{1},y_{2}]$ and no treatment was applied) then we can
construct the partition tree using the auxiliary sample and bring
the partition over to the main sample. If the main sample is smaller,
then it can be pruned back until the minimum cell has at least $c_{B}$
units. As there is not sufficient data to tune this new partition
to out-of-sample performance, it might result in slightly more blocks
than are optimal.
\end{rem}

\subsection{Zero pre-period outcomes}

If no pre-treatment outcomes exist, but there are covariates $X$,
then one alternative would be to use an unsupervised dimension reduction
technique such as principal component analysis or neural-network autoencoders
to select the blocking variables (choosing the number of dimensions
by identifying when the marginal explained variance begins to diminish).
The partition could be constructed as an evenly-distributed quantile-based
grid granular enough that the smallest cell has size $c_{B}$. This
might result in slightly more blocks than optimal.

\end{document}

%% file: insets/tbl2_coeff_controls2.tex
\newcolumntype{C}{>{\centering\arraybackslash}X}

\begin{tabularx}{\textwidth}{lCCCC}

\toprule
{Method}&{Mexico, n=100}&{Mexico, n=300}&{Sri Lanka, n=100}&{Sri Lanka, n=300} \tabularnewline
\midrule\addlinespace[1.5ex]
FPS: Random Forest&.0242458&.007467&.0294429&.0109078 \tabularnewline
Manual: 48 blocks&.0368593&.0093043&.0355722&.0115674 \tabularnewline
VS: CART&.0246845&.0079382&.0310299&.0101904 \tabularnewline
VS: Lasso + CART&.0251519&.0071399&.0285996&.0099049 \tabularnewline
\bottomrule \addlinespace[1.5ex]

\end{tabularx}

%% file: insets/tbl2_se_controls.tex
\newcolumntype{C}{>{\centering\arraybackslash}X}

\begin{tabularx}{\textwidth}{lCCCC}

\toprule
{Method}&{Mexico, n=100}&{Mexico, n=300}&{Sri Lanka, n=100}&{Sri Lanka, n=300} \tabularnewline
\midrule\addlinespace[1.5ex]
FPS: Random Forest&509.4455&268.1693&917.4573&515.8929 \tabularnewline
Manual: 48 blocks&611.7684&300.0989&964.0424&537.3345 \tabularnewline
VS: CART&525.2979&274.4434&925.6401&499.0057 \tabularnewline
VS: Lasso + CART&514.9183&264.9388&905.8749&500.7876 \tabularnewline
\bottomrule \addlinespace[1.5ex]

\end{tabularx}